\documentclass[twocolumn, apl, amsmath, amssymb, showpacs]{revtex4-2}
\usepackage{epsf}      
\usepackage{graphicx}
\usepackage{color}
\usepackage{soul}
\usepackage{gensymb}
\usepackage{sidecap}
\usepackage{amsmath}
\usepackage{mathtools}

\begin{document}

\title{Unconventional magnetic glassiness in non-centrosymmetric Sm$_7$Pd$_3$: Interplay of magnetic frustration, long-range order, and frozen domains}

\author{Ajay Kumar}
\email{ajay1@ameslab.gov}

\author{Anis Biswas}

\author{Yaroslav Mudryk}

\affiliation{Ames National Laboratory, U.S. Department of Energy, Iowa State University, Ames, Iowa 50011, USA}

\date{\today}

\begin{abstract}

We present a comprehensive investigation of the intricate spin dynamics in the non-centrosymmetric compound Sm$_7$Pd$_3$, revealing the coexistence of spin glass, domain glass, and ferromagnetic (FM) behaviors. Magnetic field-dependent measurements indicate large coercivity, suggesting ferromagnetic domain formation below the Curie temperature ($T_{\rm C} \approx 173$ K), while temperature-dependent magnetization data point to antiferromagnetic (AFM) coupling, highlighting the competition between FM and AFM interactions. Detailed ac susceptibility, isothermal remanent magnetization, and aging effect measurements demonstrate the presence of two distinct types of glassiness in the sample, and their possible origins are discussed extensively. Magnetization measurements reveal the mixing of the $J = 5/2$ ground state  of Sm$^{3+}$ with the excited $J = 7/2$ multiplet, lying 965 K above. The specific heat data show further crystalline electric field splitting of the $J = 5/2$ state into a ground-state doublet and a fourfold-degenerate excited state.

\end{abstract}
\maketitle

\section{\noindent ~Introduction}

Describing the magnetic state of a complex compound with multiple magnetically active atomic positions is often a non-trivial task. Such compounds rarely conform to classic definitions of ferromagnetic (parallel moment alignment inside magnetic domains), antiferromagnetic (antiparallel moment alignment resulting in near-zero magnetization), or canonical spin-glass (equal probability of AFM and FM configurations due to geometrical frustration of the spin system) order \cite{Cheong_NPJ_24, Khmelevskyi_CP_24, Bastien_PRB_19}. Altermagnetism is a telling example of an ordered system where the classic approach is no longer sufficient to define the magnetic state \cite{Mazin_PRX_22, Krempasky_Nature_24, Smejkal_PRX_22}, and, likely, it will not be the last one. Often magnetic systems exhibit multiple characteristic behaviors – for example, displaying FM behavior along one crystallographic direction and AFM behavior along another or combine magnetic frustration with well-established long-range order \cite{Ma_PRL_09, Castro_PRB_12, Chattopadhyay_PRB_04}. Acquiring a fundamental understanding of magnetism in such materials presents a substantial challenge, particularly when conventional approaches, such as magnetic structure determination via neutron diffraction, are not feasible due to the inherently low scattering cross section of the  constituent elements and/or the presence of complex dynamic spin fluctuations \cite{Sanchez_NC_21}. Yet to advance the current paradigm of magnetic ordering, we must explore and explain these complex examples with utmost rigor.

The non-centrosymmetric intermetallic compounds with the general formula  R$_7$T$_3$ (where R represents a rare-earth element and T denotes a transition metal) have attracted significant research interest due to their diverse range of exotic physical phenomena, including unconventional electrical transport, large coercivity with nearly zero magnetization, exchange bias, giant magnetocaloric effect (GMCE), and glassy magnetic states \cite{Barker_PRL_15, Singh_PRB_20, Sengupta_PRB_06}. For example, the lack of inversion symmetry in the crystal structure induces unconventional superconductivity in La$_7$T$_3$ (T = Ir, Rh, and Ni) via time-reversal symmetry breaking \cite{Barker_PRL_15, Singh_PRB_20, Arushi_PRB_21}. All R$_7$T$_3$ compounds crystallize in the hexagonal Th$_7$Fe$_3$-type structure ($P6_3mc$) at room temperature, where the rare-earth atoms occupy three distinct crystallographic sites (two at $6c$ and one at $2b$) \cite{Moreau_JLCM_73}. Each $6c$ site belongs to a distinct atomic plane and one of these planes is shared with $2b$ position.  This intricate structural arrangement results in competing magnetic interactions, with small energy differences between various configurations, leading to frustrated ground states.  Furthermore, the delicate interplay between the localized $4f$ moments of R atoms and the itinerant $d$ states of T atoms enhances the complexity of the magnetic and electronic structures of these compounds \cite{Gubkin_PRB_22, Mudryk_JPCM_19, Umeo_PRB_03}. 

For instance, Nd$_7$Rh$_3$ exhibits two antiferromagnetic (AFM) transitions at T$_{\rm N1}$ = 32 K and T$_{\rm N2}$ = 10 K, along with an anomalous field-induced first-order AFM-to-FM transition below T$_{\rm N2}$, exhibiting the signature of phase coexistence after the field cycling \cite{Sengupta_PRB_06}. The coexisting FM and AFM regions form a phase separated state, which is distinct from a typical spin-glass state and gives rise to a unconventional time-dependent magnetic behavior \cite{Sengupta_PRB_06}. Termed ``magnetic glass" \cite{Sengupta_PRB_06, Pal_PRB_21} to differentiate from ``spin glass", such states are common in compounds with strong magnetostructural coupling, including rare earths manganates and kinetically arrested Gd$_5$Ge$_4$ \cite{Dho_PRB_03, Chattopadhyay_PRB_04, Pal_PRB_21, Pal_PRB_23}. Similar order-order successive magnetic transitions are observed in Ho$_7$Rh$_3$: a transition from a paramagnetic state to an incommensurate AFM structure at T$_{\rm N}$ = 32 K,  followed by a spin reorientation transition around 22 K, and finally,  the emergence of a ferromagnetic component for T $\lesssim$ 9 K \cite{Gubkin_PRB_22}. Notably, short-range magnetic correlations persist in Ho$_7$Rh$_3$ well above T$_{\rm N}$ ($\sim$2T$_{\rm N}$), underscoring the presence of competing interactions even in the absence of long-range order  \cite{Gubkin_PRB_22}. A particularly intriguing example is Nd$_7$Pd$_3$, which undergoes a second-order paramagnetic-to-AFM transition at T$_{\rm N}$ = 37 K, followed by a first-order AFM-to-FM transition at T$_{\rm C}$ = 33 K, accompanied by a structural change from $P6_3mc$ to $Cmc2_1$, which is necessary to accommodate the ground state FM structure \cite{Mudryk_JPCM_19}. The close proximity of T$_{\rm N}$ and T$_{\rm C}$ in this compound underscores the fragility of AFM state \cite{Mudryk_JPCM_19}. We note that first-order transition at T$_{\rm C}$ is practically anhysteretic with giant magnetocaloric effect and, if not for a high cost of Pd, could be considered as a very promising GMCE material \cite{Mudryk_JPCM_19, Singh_JPCM_09}.

Among R$_7$Pd$_3$ compounds, our previous work highlights the remarkable magnetic behavior of Sm$_7$Pd$_3$.  Sm$_7$Pd$_3$ crystallizes in the Th$_7$Fe$_3$-type structure at room temperature and orders ferro- or ferrimagnetically below T$_{\rm C}$ $\approx$170 K \cite{Kadomatsu_JMMM_98, Biswas_AM_24}. Our previous study establishes that the crystal symmetry of Sm$_7$Pd$_3$ remains $P6_3mc$ down to 5 K; however, significant changes in lattice parameters, unit cell volume, and the $c/a$ ratio occur at T$_{\rm C}$, even though this magnetoelastic transition is of second order \cite{Biswas_AM_24}. Sm$_7$Pd$_3$ exhibits extraordinarily large coercivity, while having nearly zero saturation magnetization—an ideal combination for efficient spin-transfer and memory applications \cite{Kadomatsu_JMMM_98, Biswas_AM_24, Jungwirth_NT_16, Jungwirth_NM_22}. While the compound itself orders below room temperature, understanding the control knobs for tuning these technologically relevant properties presents an important fundamental and applied scientific problem. Our earlier study also uncovered the possible presence of antiferromagnetic interactions in the predominantly FM compound and DFT calculations showed that antiparallel moment alignment of Sm atoms from different atomic planes is energetically favourable \cite{Biswas_AM_24}. We hypothesize that they give rise to a large exchange bias effect and a crossover from positive to negative magnetization as a function of temperature in Sm$_7$Pd$_3$ \cite{Kadomatsu_JMMM_98, Biswas_AM_24}. This crossover indicates strong blocking of the magnetic moments in the sample at low temperatures due to highly frustrated spin dynamics. However, the exact role of these competing interactions in shaping its observed magnetic behavior is unknown.

In addition, the Sm-based intermetallic compounds typically show a very small energy difference between the ground state (J = 5/2), and the first (J = 7/2, 0.1293 eV) and second (J = 9/2, 0.2779 eV) excited multiplets of Sm$^{3+}$. This energy separation can be effectively altered by changing the crystal field through external perturbations, such as applied magnetic fields, pressure, or chemical substitution \cite{Wijn_PSS_76, Liu_PRB_01, Malik_PRB_79, Barla_PRB_05, Wijn_PRB_67, Pospisil_PRB_10}. As a result, the magnetic and thermodynamic properties of these compounds are governed by the degree of mixing between these multiplets, with the Van Vleck contribution playing a crucial role in their magnetic susceptibility \cite{Liu_PRB_01, Ahn_PRB_07}. Therefore, an accurate description of such mixing is necessary for a fundamental  understanding of magnetic behavior in Sm compounds.

In this work, we present a comprehensive investigation into the dynamic magnetic response of Sm$_7$Pd$_3$ and establish that it is a novel magnetic system at the crossroads of unconventional spin-glass, domain-glass, and ferromagnetic behavior. We identify two distinct types of glassiness: one exhibiting typical spin-glass behaviour born from the interplay between dominant FM and residual AFM interactions and another showing pronounced field-induced spin relaxation reminiscent of phase-separated states observed in Nd$_7$Rh$_3$, originating from domain locking. The underlying mechanisms have been elucidated through detailed ac and dc susceptibility measurements under various protocols. Furthermore, we analyze the mixing of the ground-state J = 5/2 and the excited J = 7/2 multiplets of Sm$^{3+}$ and their splitting under the influence of crystalline electric field using both magnetization and specific heat measurements.

\begin{figure*}  
\centering
\includegraphics[width=0.75\textwidth]{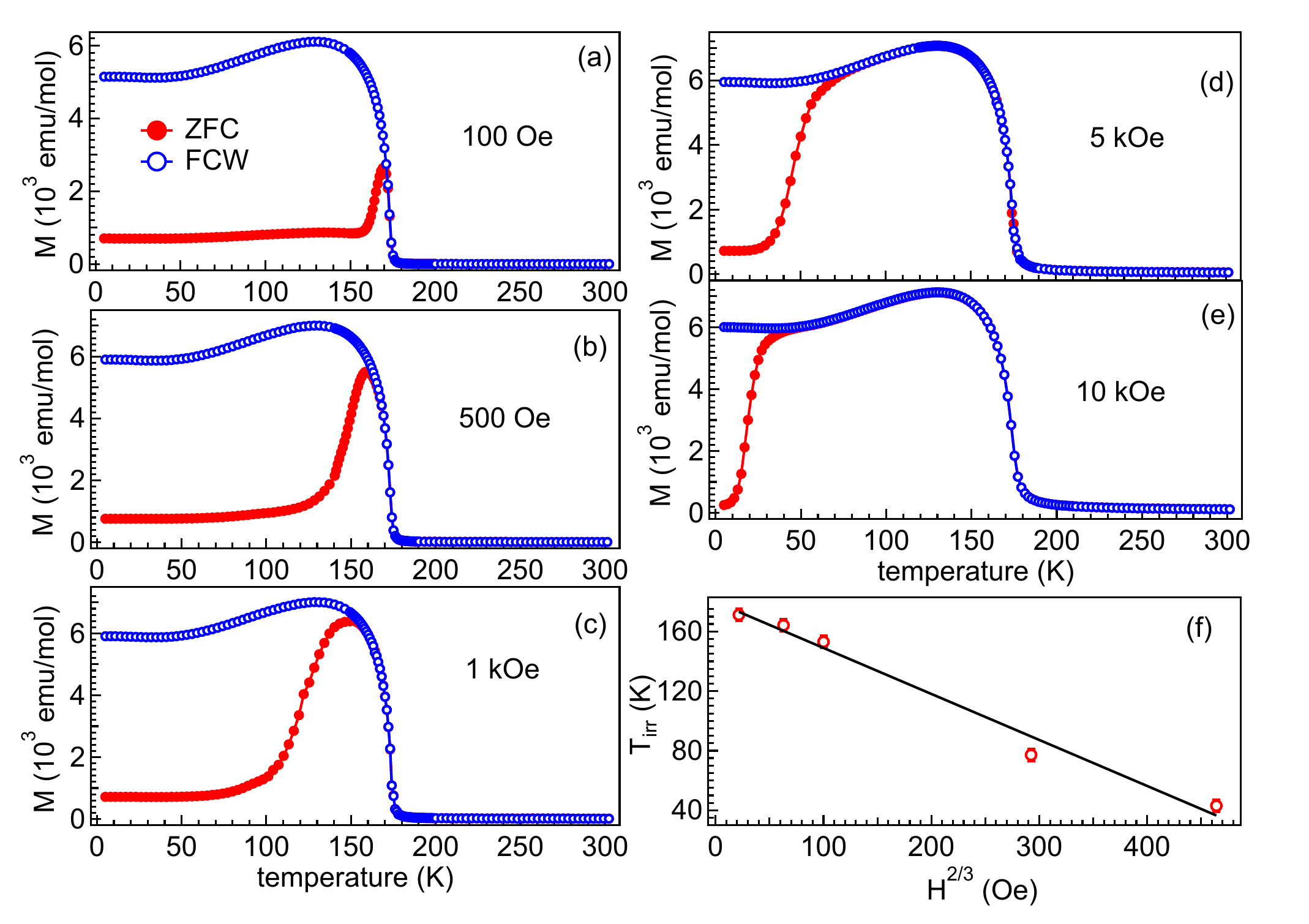}
\caption {(a--e) The temperature dependent magnetization in zero-field-cooled (ZFC) and field-cooled warming (FCW) modes at different applied magnetic fields. (f) The field dependence of the irreversibility temperature plotted as T$_{\rm irr}$ vs. H$^{2/3}$. The solid black line represents the linear fit of the data.} 
\label{Fig1_ZFC_FC}
\end{figure*}

\section{\noindent ~Experimental}

The polycrystalline Sm$_7$Pd$_3$ sample was prepared by arc melting stoichiometric amounts of pure Sm (99.95\%), provided by the Materials Preparation Center of Ames National Laboratory, and Pd (99.99\%), supplied by Vale Acton Refinery, London, UK. A Zr getter was melted prior to the sample melting to absorb the residual oxygen in the chamber. The sample was flipped and remelted several times to ensure homogeneity. Two samples: one melted four times, and another melted six times, were prepared to ensure reproducibility. No difference in either XRD pattern or magnetization data was observed between the two samples.

The room-temperature x-ray diffraction (XRD) pattern was recorded using a Rigaku TTRAX powder diffractometer equipped with a rotating anode Mo K$\alpha$ source \cite{Holm_RSI_04}. The Rietveld refinement of the XRD data was performed using  the FullProf software, which confirm an almost phase-pure (with a $<$1\% unknown phase) Th$_7$Fe$_3$-type hexagonal crystal structure of the compound [see Fig. 1(a, b) of \cite{SI}], consistent with the reported literature \cite{Biswas_AM_24, Kadomatsu_JMMM_98}. Temperature and magnetic field (both dc and ac) dependent magnetic measurements were conducted on a randomly oriented polycrystalline sample using a superconducting quantum interference device (SQUID) from Quantum Design, USA (model MPMS XL-7). The same SQUID magnetometer was used to perform relaxation measurements (isothermal remanent and aging effects) under different protocols. The details of the various measurement protocols are provided along with their respective discussions. Temperature and magnetic field dependent specific heat measurements were performed using the conventional 2$\tau$ relaxation technique in a physical property measurement system (PPMS) from Quantum Design, Inc.

\section{\noindent ~Results and discussion}

The temperature-dependent magnetization measurements have been performed under zero-field-cooled (ZFC) and field-cooled warming (FCW) conditions at different magnetic fields, as shown in Figs. \ref{Fig1_ZFC_FC}(a-e). A clear PM to FM-like transition has been observed at $T_{\rm C} = 173 \pm 1$ K, which is slightly higher than those reported in \cite{Kadomatsu_JMMM_98} (167 K) and \cite{Biswas_AM_24} (169 K). More importantly, a large bifurcation between the ZFC and FCW curves is observed below $T_{\rm C}$, and the irreversibility temperature ($T_{\text{irr}}$) shifts to lower values with an increase in the magnetic field [see Figs. \ref{Fig1_ZFC_FC}(a-e)]. This is commonly associated with either blocking of the spins due to magnetic frustration or random locking of domain orientations during zero-field cooling, where the higher applied magnetic fields align the moments even well within the blocked regime \cite{Williams_PRB_03, Marcano_PRB_07, Kumar_PRB_24}. Interestingly, a decrease in magnetization has been also observed in the FCW curves [where the sample was cooled after aligning the moments in the PM (unblocked) state] below $\sim$130 K for all studied magnetic fields [see Figs. \ref{Fig1_ZFC_FC}(a-e)], indicating the presence of weak but finite antiferromagnetic interactions in the sample. The presence of mixed FM and AFM interactions can give rise to magnetic frustration, which may contribute to the observed large bifurcation in the ZFC-FCW curves at low temperatures.    

Another factor to consider is a strong magnetic anisotropy of the sample, predicted by earlier DFT calculations \cite{Biswas_AM_24}. Therefore, we analyze the field dependence of $T_{\text{irr}}$ using the following power-law behavior \cite{Kotliar_PRL_84, Kaul_JPCM_98, Shand_PRB_10}:

\begin{eqnarray}
T_{\rm irr}(H)=T_{\rm irr}(0)[1-AH^p]
\label{Tirr}
\end{eqnarray}

where $T_{\text{irr}}(0)$ is the irreversibility temperature in the zero-field regime, $A$ is the constant, and the power exponent $p$ represents the measure of the strength of anisotropy in the system with respect to the external magnetic field. In the strongly anisotropic regime, the value of $p$ is 2/3 in accordance with the de Almeida-Thouless (A-T) line for Ising spins \cite{Almeida_JPCG_78}. On the other hand, weakly anisotropic Heisenberg spin systems are defined by the Gabay-Toulouse (G-T) line with $p \approx 2$ \cite{Gabay_PRL_81}. We observed a significant deviation in $T_{\text{irr}}$ from the G-T line, as shown in Fig. 2(c) of \cite{SI}. However, good agreement with the A-T line is observed up to $H = 10$ kOe, as evident from the linear relationship between $T_{\text{irr}}$ and $H^{2/3}$, shown in Fig. \ref{Fig1_ZFC_FC}(f). Fitting of the $T_{\text{irr}}(H)$ data using the A-T line, represented by the solid black line in Fig. \ref{Fig1_ZFC_FC}(f), gives $T_{\text{irr}}(0) = 180(6)$ K and $A = 0.0017(1)$ Oe$^{-2/3}$. The good agreement of the $T_{\text{irr}}(H)$ data with the A-T line indicates the presence of strong anisotropy in the system.

\begin{figure} 
\includegraphics[width=0.5\textwidth]{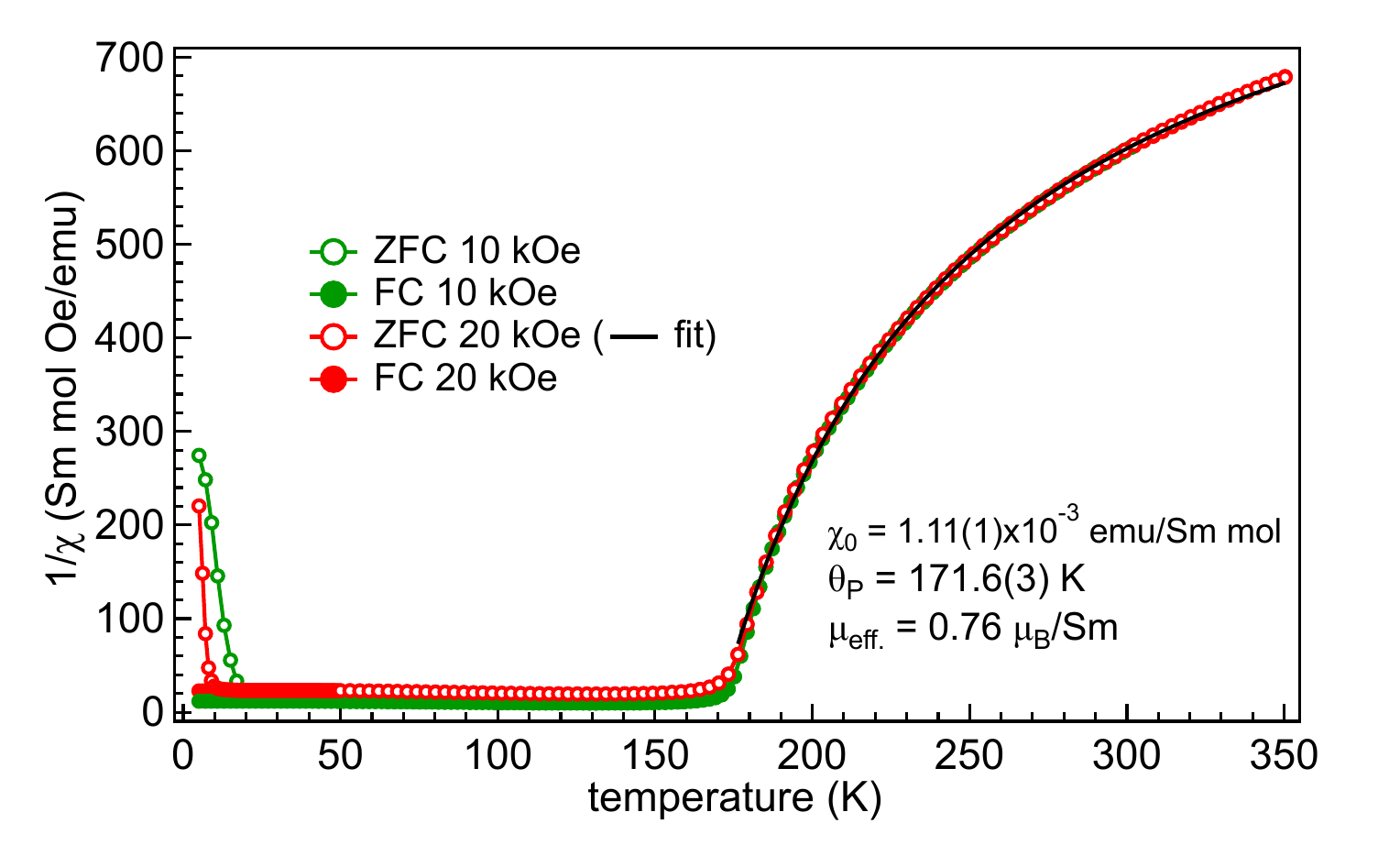}
\caption {The temperature dependent inverse magnetic susceptibility of Sm$_7$Pd$_3$ at 10 and 20~kOe fields in ZFC and FCW modes. Solid black curve represents the best fit of the 20~kOe ZFC data using the modified Curie-Weiss law. }
\label{Fig2_MT_CW}
\end{figure}

\begin{figure}
\centering
\includegraphics[width=3.5in]{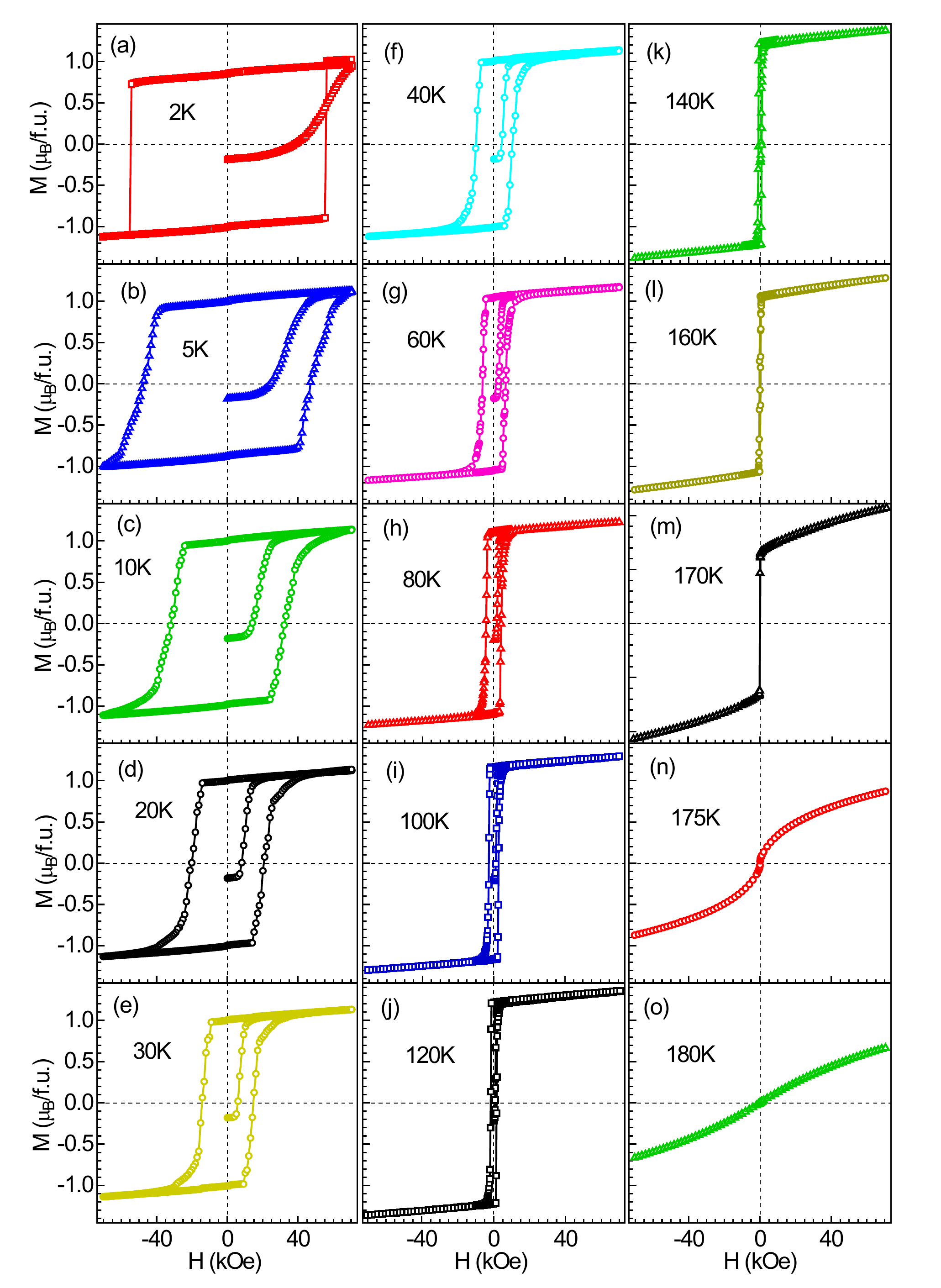}
\caption {(a--o) The ZFC M-H curves of Sm$_7$Pd$_3$ between -70 and 70 kOe magnetic fields at different temperatures. } 
\label{Fig3_MH}
\end{figure}

\begin{figure}
\centering
\includegraphics[width=3.5in]{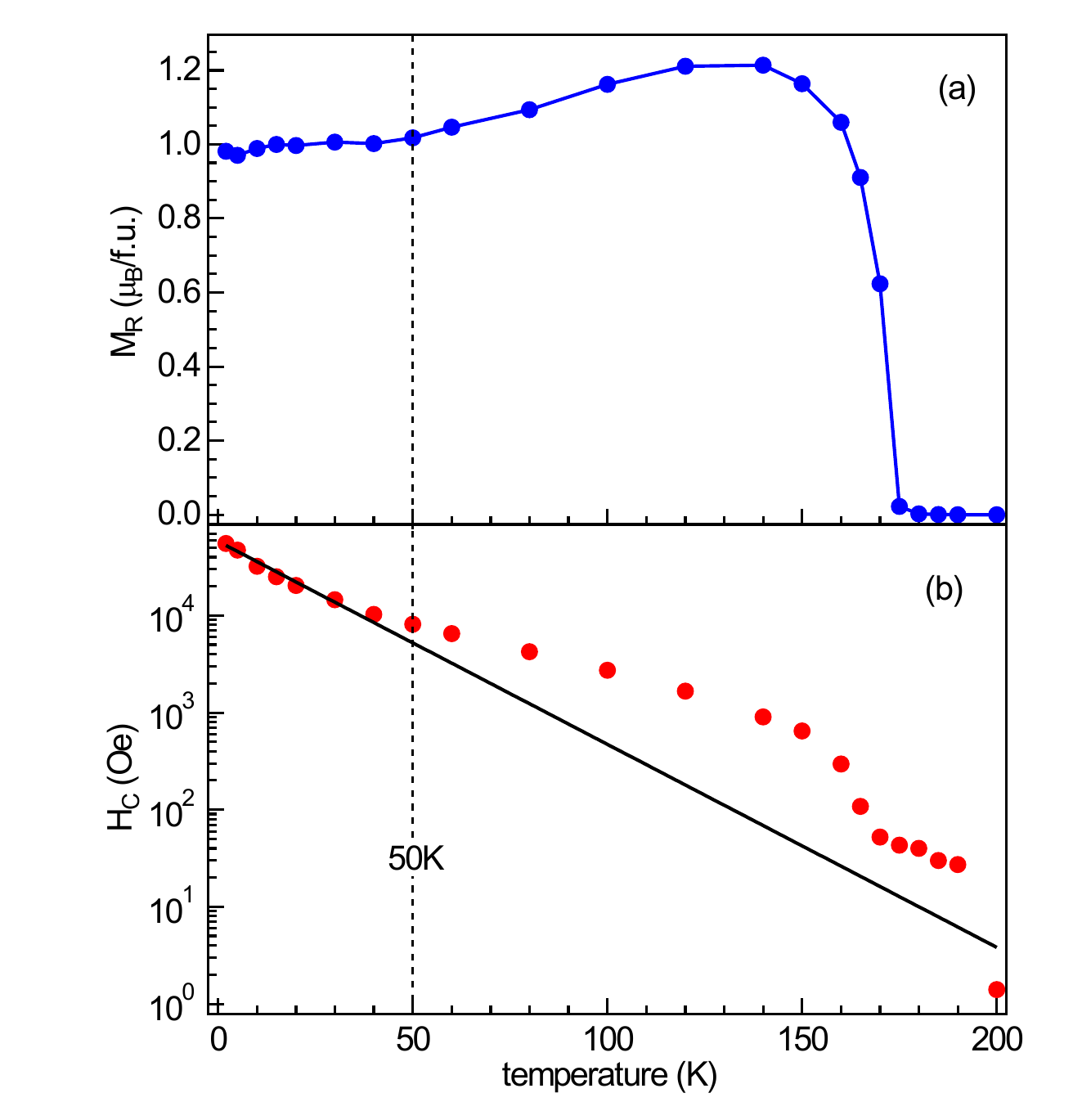}
\caption {The temperature dependent (a) remanent magnetization (M$_{\rm R}$) and (b) coercive field (H$_{\rm C}$) (on the semi logarithmic scale) of Sm$_7$Pd$_3$. The solid black line in (b) represents the best fit of the data using eq.  \ref{Hc}. Vertical  dashed line shows the deviation in the M$_{\rm R}$ and H$_{\rm C}$ above 50~K. } 
\label{Fig4_coercivity}
\end{figure}

In Fig. \ref{Fig2_MT_CW}, we show the temperature-dependent inverse magnetic susceptibility ($\chi^{-1} = H/M$) at H = 10 and 20 kOe [Fig. 2(a) of \cite{SI}] in both ZFC and FCW modes. A field-independent strong non-linearity in the curves can be clearly observed in the paramagnetic state, indicating the mixing of the ground state $J = 5/2$ multiplet of Sm$^{3+}$ with the excited $J = 7/2$ level, resulting in the non-zero second-order Zeeman splitting \cite{Barla_PRB_05, Ahn_PRB_07, Malik_PRB_79, Wijn_PRB_67, Pospisil_PRB_10}. We fit the 20 kOe ZFC data using the modified Curie-Weiss law given as \cite{Ahn_PRB_07, Yuhasz_PRB_05}:

\begin{eqnarray}
\chi = \chi_0 + \frac{C}{T - \theta_p} = \left( \frac{N_A}{k_B} \right) \left( \alpha_J \mu_B^2 + \frac{\mu_{\rm eff}^2}{3(T - \theta_P)} \right), 
\label{CW}
\end{eqnarray}

where $\chi_0$ is the temperature-independent Van Vleck susceptibility, C and $\theta_P$ are the Curie-Weiss constant and Curie temperature, respectively. N$_A$, k$_B$, $\mu_B$, and $\mu_{\rm eff}$ are Avogadro's number, Boltzmann constant, Bohr magneton, and effective magnetic moment, respectively. The constant $\alpha_J$ is expressed as 

\begin{align}
\alpha_J &= k_B \left\{ 
\frac{\left[ J^2 - (L - S)^2 \right] \left[ (L + S + 1)^2 - J^2 \right]}{6J(2J + 1) \left( E_{J-1} - E_J \right)} \right.  + \notag \\ 
&\quad \left. \frac{\left[ (J+1)^2 - (L - S)^2 \right] \left[ (L + S + 1)^2 - (J+1)^2 \right]}{6(J+1)(2J + 1) \left( E_{J+1} - E_J \right)} \right\} \notag \\ 
&= \frac{20}{7\Delta}
\end{align}

where $\Delta$ is the energy separation between J = 5/2 and 7/2 states. The best fit, shown by the solid black curve in Fig. \ref{Fig2_MT_CW}, yields $\chi_0$ = 1.11(1)$\times$10$^{-3}$ emu/Sm mol Oe, $\theta_P$ = 171.6(3)~K, and C = 0.0718(5)~emu K/Sm mol Oe, which gives $\mu_{\rm eff}$ = 0.76 $\mu_B$/Sm and $\Delta$ = 965~K. The calculated value of $\theta_P$ is in good agreement with the T$_{\rm C}$ estimated from the dM/dT curves (173~K). The separation between J = 5/2 and 7/2 states is smaller than the estimated value of $\sim$1500~K for the free Sm$^{3+}$ ion \cite{Vleck_book_32}, but is in good agreement with that observed for several other Sm-based compounds \cite{Ahn_PRB_07, Yuhasz_PRB_05, Stewart_PRB_72}, indicating the strong effect of the crystal field in Sm$_7$Pd$_3$. This is also evident from the observed lower value of the effective magnetic moment (0.76 $\mu_B$/Sm) as compared to the free Sm$^{3+}$ ion (0.84$\mu_B$).\par

The reduction in the ZFC magnetization or the bifurcation in the ZFC-FCW curves at low temperatures can arise either due to the onset of a spin-glass (SG) like disordered magnetic state \cite{Gabay_PRB_96, Yeshurun_PRB_81} or to the inherent magnetic hardness of the FM system \cite{Read_JMMM_84}. To distinguish this, field-dependent magnetization (M-H) measurements were conducted at various temperatures, as shown in Figs. \ref{Fig3_MH}(a-o). The sample was heated to room temperature after recording each successive M-H curve to eliminate remanence. Notably, at 2 K, a giant coercivity ($H_{\rm C}$) of around 55 kOe, accompanied by a step-like jump in magnetization, is observed, qualitatively replicating the results obtained in our previous study \cite{Biswas_AM_24}. Note that the virgin magnetization curve at 2 K is not fully saturated even up to 70 kOe [see Fig. 3(a)], and the value of $H_{\rm C}$ can be further increased by applying higher magnetic fields \cite{Biswas_AM_24}. A small value of $\sim 0.15 \, \mu_B/\text{Sm}$ has been observed at 70 kOe (at 2 K), which is significantly lower than the theoretical saturation moment ($g_J J$) of 0.71 $\mu_B$ for the free Sm$^{3+}$ ion, indicating the strong effect of the crystal field and/or canting of the moments. The magnetic state comprising large coercivity along with nearly zero saturation magnetization is highly desirable for efficient spintronics and memory device applications \cite{Jungwirth_NT_16, Jungwirth_NM_22}, and, therefore, demands further investigation. 

A negative magnetization is detected at the beginning of the virgin isotherms, possibly due to the presence of a nonzero remanent field in the SQUID coils during nominally ``zero-field" cooling. Below T$_{\rm C}$, the coercivity, $H_{\rm C}$, as well as the sharpness of the M-H loop, decreases, whereas the remanent magnetization ($M_{\rm R}$) remains almost invariant with increasing temperature [see Figs. \ref{Fig3_MH}(a-m)]. In fact, the $M_{\rm R}$ shows an enhancement above $\sim$50 K, and to quantitatively understand this, the temperature dependence of $M_{\rm R}$ and $H_{\rm C}$ have been plotted in Figs. \ref{Fig4_coercivity}(a) and \ref{Fig4_coercivity}(b), respectively. The remanent magnetization remains nearly invariant up to around 50 K and then starts increasing with temperature [marked by the dashed line in Fig. \ref{Fig4_coercivity}(a)], followed by a sharp reduction at $T_{\rm C}$. This behavior is analogous to the FCW $M(T)$ curves discussed above [see Figs. \ref{Fig1_ZFC_FC}(a-f) and Fig. 3 of \cite{SI} for more clarity], confirming the presence of AFM coupling in the sample at low temperatures, which becomes less pronounced with an increase in temperature above 50 K. Furthermore, the coercivity decreases abruptly with increasing temperature, which for a mixed FM-AFM system, can be described by the following relation \cite{Moutis_PRB_01}:

\begin{eqnarray} 
H_C = H_{\rm C}(0)\exp(-T/T_1), 
\label{Hc}
\end{eqnarray}

where H$_{\rm C}$(0) is the coercivity at 0~K and T$_1$ is a constant. Interestingly, we observe an increase in the coercivity from the exponential decay behaviour, defined by Eq. \ref{Hc}, above around 50~K, as depicted by solid black line in Fig. \ref{Fig4_coercivity}(b) for H$_{\rm C}$(0) = 58(2)~kOe and T$_1$ =  21(1)~K. This further confirms the enhancement in the fraction of FM interactions in the sample above this temperature.

\begin{figure} [h]
\includegraphics[width=3.5in]{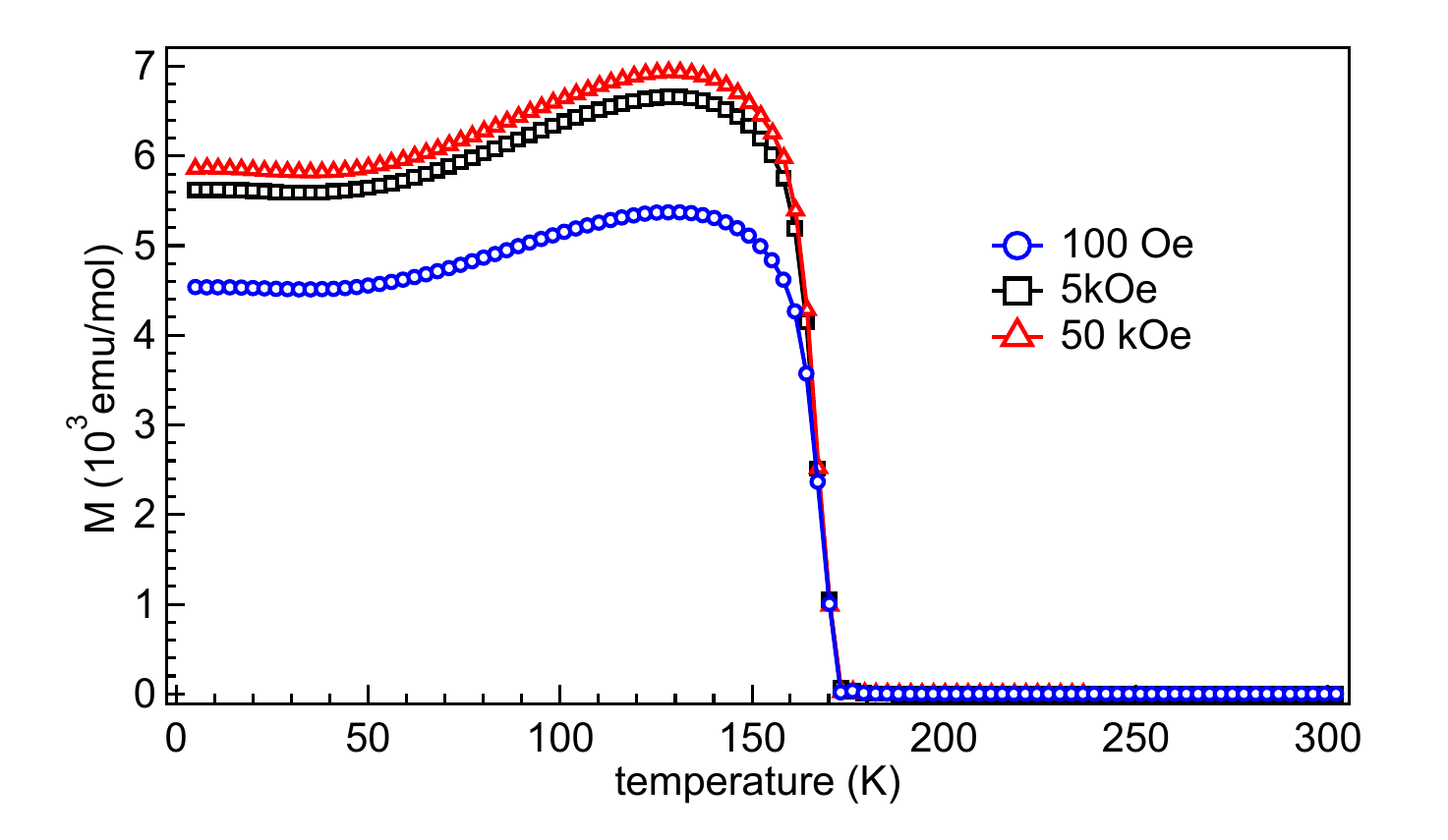}
\caption {The thermoremanent magnetization (TRM) measurements performed on Sm$_7$Pd$_3$ at different cooling fields.} 
\label{Fig5_TRM}
\end{figure}

The temperature-induced change in the $M_{\rm R}$ is further probed by the thermoremanent magnetization (TRM) measurements performed at different magnetic fields, as shown in Fig. \ref{Fig5_TRM}. To record the TRM data, we first cool the sample from 300 K (PM state) to 2 K in the presence of an applied magnetic field, then remove the field at 2 K and record the magnetization as a function of temperature in zero external field. A clear enhancement in magnetization is observed above $\sim$50 K in Fig. \ref{Fig5_TRM} for all cooling fields. An increase in the sample’s magnetization, even in the absence of any external magnetic field, confirms the presence of AFM coupling in the sample, which increases for $T \lesssim$ 50 K and weakens with an increase in temperature. It is interesting that this process is continuous, without a well-defined spin-reorientation transition, and one may reasonably assume that the rapid but continuous change in $c/a$ ratio below the $T_{\rm C}$ observed in \cite{Biswas_AM_24} affects the relative strength, and, possibly, sign of exchange interactions between Sm atoms located on different planes, which are stacked along the $c$-axis.

\begin{figure}
\centering
\includegraphics[width=3.5in]{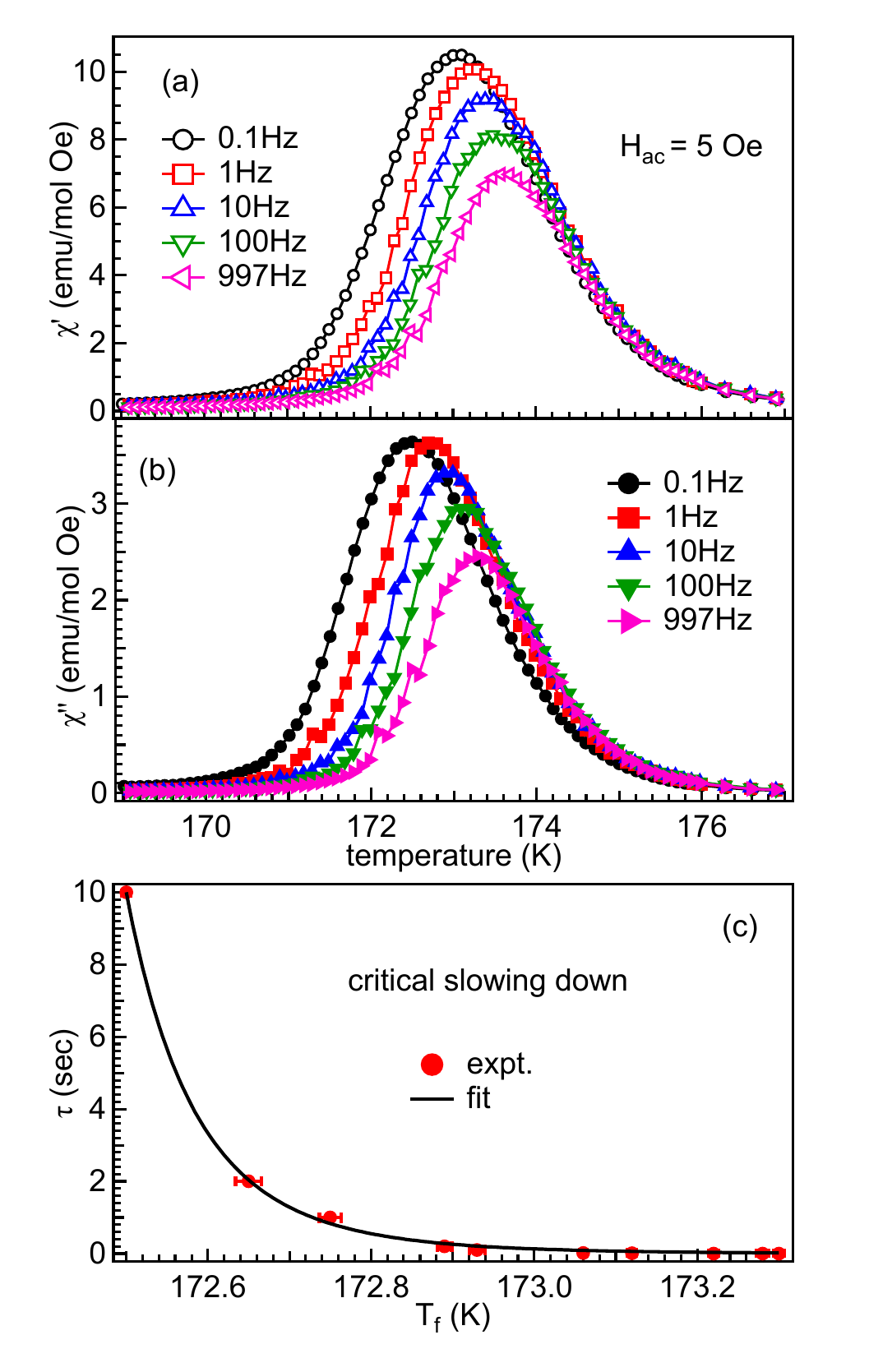}
\caption {The temperature dependent (a) real and (b) imaginary components of the ac susceptibility of Sm$_7$Pd$_3$ at different frequencies of 5~Oe ac field. (c) The frequency dependence of the freezing temperature, plotted as $\tau$=1/$f$ vs. T$_f$. The solid black curve represents the best using the critical slowing down model.} 
\label{Fig6_ac}
\end{figure}

The presence of mixed FM-AFM interactions often results in magnetic frustration, which leads to SG-like behavior in the samples \cite{Feng_PRB_06, Hiroi_PRB_09, Guan_PRM_22}. To probe this, we performed ac susceptibility measurements on Sm$_7$Pd$_3$ in the vicinity of the magnetic transition using ac frequencies from 0.1 to 997 Hz. The temperature dependent real ($\chi'$) and imaginary ($\chi''$) components of the ac susceptibility, measured at $H_{ac} = 5$ Oe, are presented in Figs. \ref{Fig6_ac}(a) and (b), respectively. A clear peak in both $\chi'$- and $\chi''$-T curves around 173 K confirms the presence of spin blocking. A reduction in magnitude and a shift of the peak position toward higher temperatures have been observed with an increase in the frequency for both the real and imaginary susceptibilities, indicating the spin-glass behavior \cite{Binder_RMP_86}. In order to quantitatively characterize this SG behavior, we calculate the shift in the peak position per decade change in the frequency, defined by the Mydosh parameter as $\partial T_f = \Delta T_f/T_f \Delta (\log_{10} \omega)$, where $\Delta T_f$ is the shift in freezing temperature ($T_f$) and $\omega = 2\pi f$ is the angular frequency \cite{Mydosh_Book_93}. It has been established that the value of $\partial T_f$ lies between 0.005 -- 0.01 for canonical spin glass systems, between $\sim$0.03 -- 0.08 for cluster glasses, and $\partial T_f > 0.2$ for superparamagnetic systems \cite{Mydosh_Book_93, Giot_PRB_08, Kumar_PRB_20, Anand_PRB_12}. The $\partial T_f \to 0$ for conventional long-range magnetic ordering (AFM/FM), where the moments align almost instantly in the direction of the applied ac magnetic field. The calculated value of $\partial T_f$ for Sm$_7$Pd$_3$ is 0.0011, which puts it in the range between canonical spin glass and conventional long-range magnetic ordering. This means that even though magnetic glassiness is present in the sample, as evident from the frequency dispersion of $\chi'(T)$ and $\chi''(T)$ curves, long-range magnetic ordering coexists with the magnetic frustration.

To further understand magnetic frustration in Sm$_7$Pd$_3$, the frequency dependence ($\tau$ = 1/$f$) of the peak position of $\chi^{\prime \prime}$ (T) curves is fitted using the critical slowing down model, given as \cite{Mydosh_Book_93, Hohenberg_RMP_12}

\begin{eqnarray} 
\tau = \tau_0\left(\frac{T_f}{T_{\rm SG}}-1\right)^{z\nu}
\label{critical}
\end{eqnarray}

where $\tau_0$ denotes the characteristic relaxation time associated with an individual spin flip, $T_{\rm SG}$ is the freezing temperature at $f = 0$ Hz, and $z\nu$ is the dynamical critical exponent \cite{Mydosh_Book_93, Hohenberg_RMP_12}. The black solid curve in Fig. \ref{Fig6_ac}(c) represents the best fit of the data using the critical slowing down model, which gives $T_{\rm SG} = 171.9(3)$ K, $z\nu = 7.3(2)$, and $\tau_0 = 1.42 \times 10^{-17}$ sec. The value of $\tau_0$ lies between $10^{-12}$ -- $10^{-14}$ sec for non-interacting canonical spin glass systems and between $10^{-7}$ -- $10^{-10}$ sec for cluster-glass systems with a finite interaction between the spins \cite{Kumar_PRB_20, Anand_PRB_12}. In the present case, the characteristic relaxation time, $\tau_0$, is even smaller than that of canonical spin-glass systems, which confirms the presence of coexisting long-ranged FM (where $\tau_0 \to 0$) and short-range SG-like interactions in the sample. Similar unconventional ferromagnetism produced by coexisting FM-AFM interactions has been observed in several other compounds \cite{Hiroi_PRB_09, Ceccarelli_PRB_11, Feng_PRB_06, Anand_PRB_12, Hertz_PRE_99, Wu_PRB_03}, where the strength of AFM is weaker compared to the FM coupling, yet sufficient to introduce frustration in the magnetic moments.

The frequency dispersion in the $\chi'$ and $\chi''$ is observed predominantly near the magnetic transition (see Fig. 4 of \cite{SI}). Therefore, to further understand the spin dynamics and the distribution of the relaxation time of the glassy/frozen spins in the sample, the ac susceptibility is measured at selected temperatures with varying frequencies, as depicted in Figs. \ref{Fig7_Cole}(a–e), in the form of a Cole-Cole plot ($\chi''$ vs. $\chi'$). The Cole-Cole plot exhibits a perfect semicircle if all the spins have the same relaxation time, whereas a distortion in the semicircle indicates a distribution of the relaxation times of the spin freezing process \cite{Mydosh_Book_93}. We fit the $\chi''$ ($\chi'$) data at different temperatures using the following relation \cite{Mydosh_Book_93, Anand_JPCM_15}:

\begin{figure}
\centering
\includegraphics[width=3.5in]{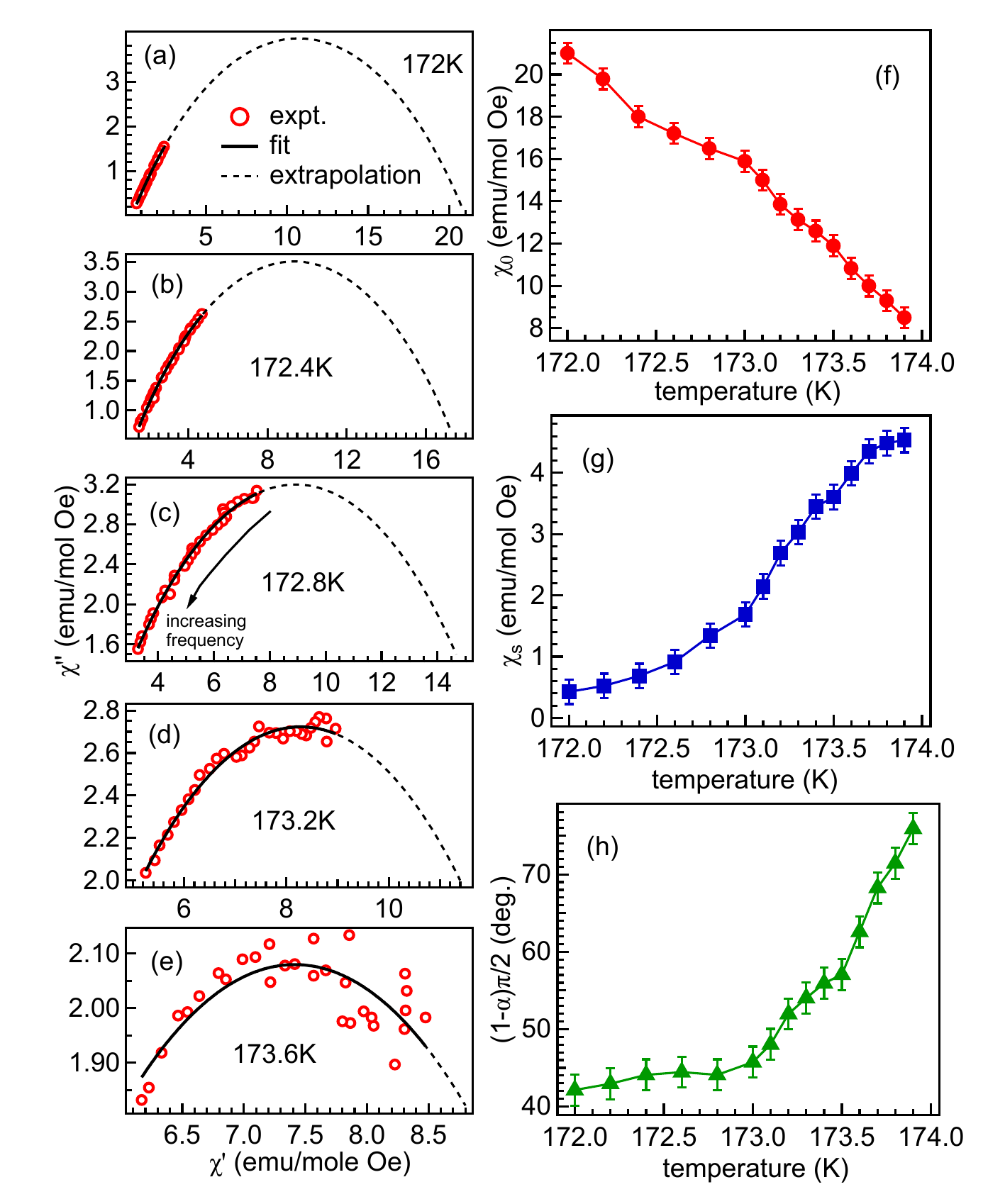}
\caption {(a--e) Cole-Cole plots ($\chi ^{\prime\prime}$ vs. $\chi ^{\prime}$) obtained from ac susceptibility data at the selected temperatures. The solid black curves represent the best fit using eq. \ref{cole} and the dashed curves are the extrapolation of the fitted curves. (f--h) The temperature dependence of the isothermal susceptibility ($\chi_0$), adiabatic susceptibility ($\chi_s$), shape parameter $\alpha$, respectively. } 
\label{Fig7_Cole}
\end{figure}

\begin{align}
\chi^{\prime\prime} &= \frac{\chi_s - \chi_0}{2{\rm tan}[(1 - \alpha)\frac{\pi}{2}]} + \notag \\
&\quad \sqrt{(\chi^\prime - \chi_s)(\chi_0 - \chi^\prime) + \frac{(\chi_0 - \chi_s)^2}{4{\rm tan}^2[(1 - \alpha)\frac{\pi}{2}]}} 
\label{cole}
\end{align}

where $\chi_0$ and $\chi_s$ are the isothermal and adiabatic susceptibilities, which represent the value of $\chi'$ in the low and high-frequency limits, respectively. The parameter $\alpha$ is a measure of the distribution of the relaxation time, where $\alpha = 0$ represents a single relaxation time (perfect semicircle) and $\alpha = 1$ indicates a wide distribution of relaxation times (flattened semicircle). The best fit of the $\chi''$ ($\chi'$) data using the above equation for selected representative temperatures is shown by the solid black curves in Figs. \ref{Fig7_Cole}(a-e), where the dashed curves represent the extrapolation of the fit. The temperature dependences of $\chi_0$, $\chi_s$, and (1-$\alpha)\pi/2$ are shown in Figs. \ref{Fig7_Cole}(f-h), respectively. The value of $\chi_0$, representing the static susceptibility, decreases with temperature [see Fig. \ref{Fig7_Cole}(f)] across the transition, analogous to the dc magnetic susceptibility, discussed above. In contrast, $\chi_s$ increases with temperature due to the release of frozen moments, enabling them to align in the direction of the external field, even at higher frequencies. Further, the value of (1-$\alpha)\pi/2$ lies around 40 -- 45$^0$ below $\sim 173$ K and then approaches 90° ($\alpha \to 0$) with further increase in temperature. A significant nonzero value of $\alpha$ in the frozen state shows the wide distribution of the relaxation time for a variety of frozen magnetic moments in Sm$_7$Pd$_3$. Such wide distribution reflects a potentially complex frozen magnetic structure that comes from three inequivalent Sm sites arranged in the non-centrosymmetric crystal structure. A rapid change in $\alpha$ around 173 K [see Fig. \ref{Fig7_Cole}(h)] indicates the onset of the PM state.

To further understand the nature of the frustration in the sample, we performed isothermal remanent magnetization (IRM) measurements at different temperatures. To record the IRM data, we first cool the sample from 300 K (PM state) to a desired temperature in the glassy region in the presence of an applied magnetic field, wait for a certain time ($t_w$), and then remove the magnetic field and record the magnetization as a function of time. Figures \ref{Fig8_IRM}(a-e) show the IRM data recorded at some representative temperatures across the magnetic transition with a cooling field of 100 Oe and $t_w = 100$ sec. A clear magnetization decay is observed, in many cases even up to one hour, indicating the presence of the arrested/frustrated spin dynamics of the sample \cite{Chattopadhyay_PRB_04, Anand_PRB_12, Kumar_PRB_20}. Note that the relaxation of the magnetic moments is observed prominently near the magnetic transition, and this effect weakens towards both the lower and higher temperatures, i.e., in the deep FM and PM states, respectively [see Figs. \ref{Fig8_IRM}(a-e)]. This is expected for a FM state with a finite remanence (no relaxation in the ordered domain state) and a PM state having free moments (quick relaxation) [see Fig. 5(a, b) of \cite{SI} for IRM data at 5 and 300 K]. The time evolution of the magnetization for the coexisting spin glass and FM interactions can be expressed well using the stretched exponent model, given as \cite{Johnston_PRB_06, Anand_PRB_12}:

\begin{figure} [h]
\centering
\includegraphics[width=3.5in]{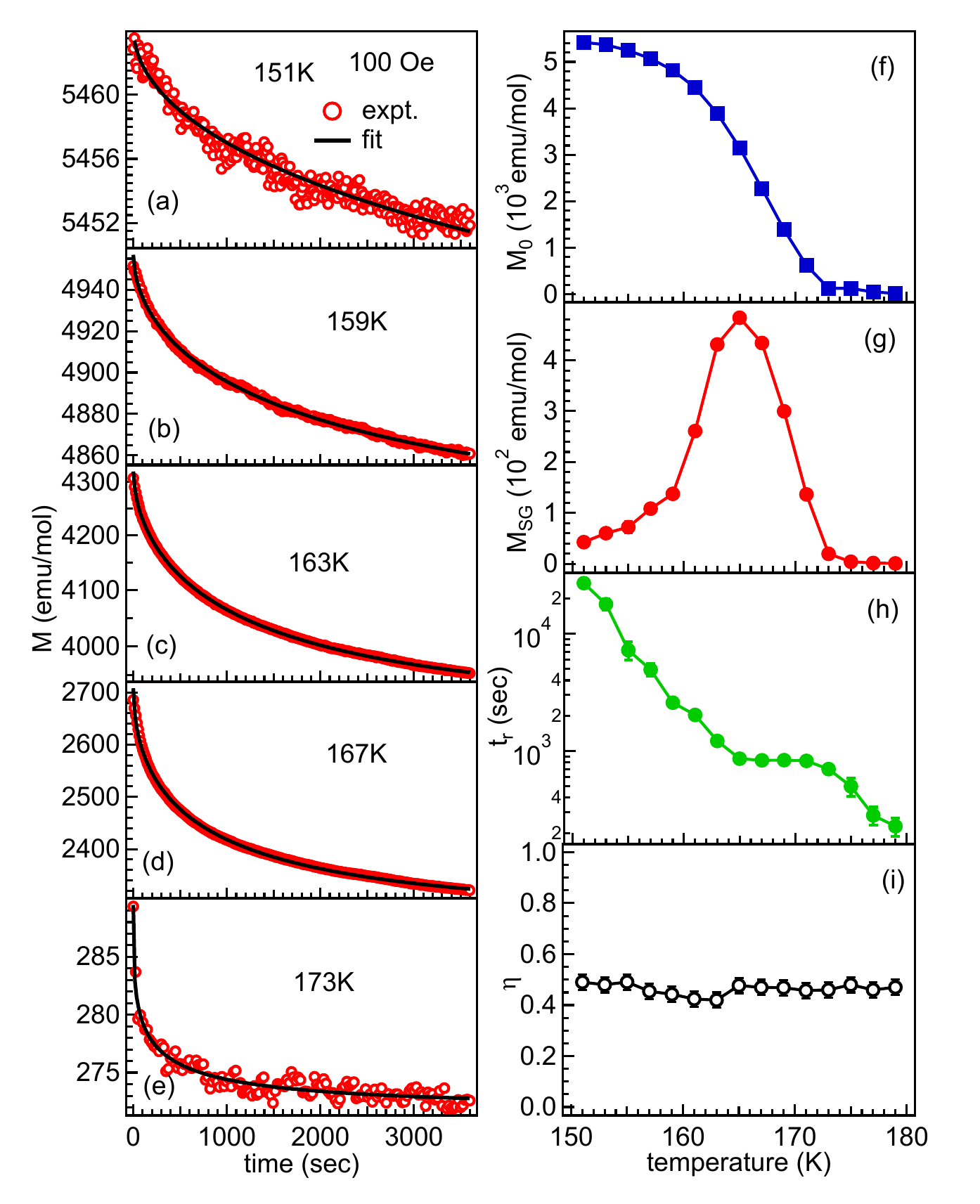}
\caption {(a--e) The field-cooled isothermal remanent magnetization (IRM) data at different temperatures for 100 Oe field and 100 sec waiting time. The solid black curves represent the best fit using stretched exponent model. The temperature dependence of the (f) FM moment, (g) spin-glass moment, (h) relaxation time, and (i) power exponent, $\eta$. } 
\label{Fig8_IRM}
\end{figure}

\begin{eqnarray} 
M(t) = M_0\pm M_{\rm SG} \times {\rm exp} \left[\left(\frac{-t}{t_r}\right)^{(1-\eta)} \right],
\label{eIRM}
\end{eqnarray}

where $M_0$ and $M_{\rm SG}$ represent the ferromagnetic and spin-glass components of the magnetic moment, respectively, and $t_r$ and $\eta$ denote the mean relaxation time and relaxation rate. The minus and plus signs correspond to the model for the IRM and the aging effect (discussed below), respectively. The best fit to the IRM data using the stretched exponent model is shown by the solid black curves in Figs.~\ref{Fig8_IRM}(a--e), with the corresponding fitting parameters presented in Figs.~\ref{Fig8_IRM}(f--i) as a function of temperature. \par

The temperature dependence of $M_0$ exhibits a typical M–T behavior for a FM system near the FM–PM transition [see Fig. \ref{Fig8_IRM}(f)], confirming the presence of the finite FM component in the sample. However, the transition temperature derived from the $\frac{dM_0}{dT}$ curve is around 165 K, which is lower than the value obtained from the dc and ac magnetic susceptibility measurements ($\sim$ 173 K) presented above. Note that the time evolution of IRM in this temperature range combines the effects of both the frustrated behavior of the magnetic lattice and the relaxation of field-induced metastable states, commonly referred to as a ``magnetic glass'' state \cite{Sengupta_PRB_06, Pal_PRB_21, Pal_PRB_23, Samanta_PRB_14}, which will be discussed in more detail below. Therefore, we propose that this shift is associated with the field-induced decrease in the magnetic glass transition temperature of the sample. Interestingly, the glassy magnetization, $M_{\rm SG}$, initially increases with temperature, exhibiting a pronounced maximum at $\sim 165$ K, then decreases upon further heating, as shown in Fig. \ref{Fig8_IRM}(g) [see Fig. 6(a, b) of \cite{SI} for a direct comparison of $M_{\rm IRM}(T)$ data]. The presence of both $M_0$ and $M_{\rm SG}$ components further confirms coexistence of FM and glassy states in the sample. The relaxation time $t_r(T)$, plotted in Fig. \ref{Fig8_IRM}(h), decreases exponentially (note the logarithmic scale on the y-axis) with increasing temperature. A notable hump in the $t_r(T)$ curve is observed around 173 K [see Fig. \ref{Fig8_IRM}(h)], where the magnetic glass state is already diminished, indicating the dominance of spin-glass behavior in this temperature regime. The exponent $\eta$ remains close to $\sim 0.5$ [Fig. \ref{Fig8_IRM}(i)], suggesting a similar relaxation mechanism in the sample across this temperature range.

\begin{figure}
\includegraphics[width=3.5in]{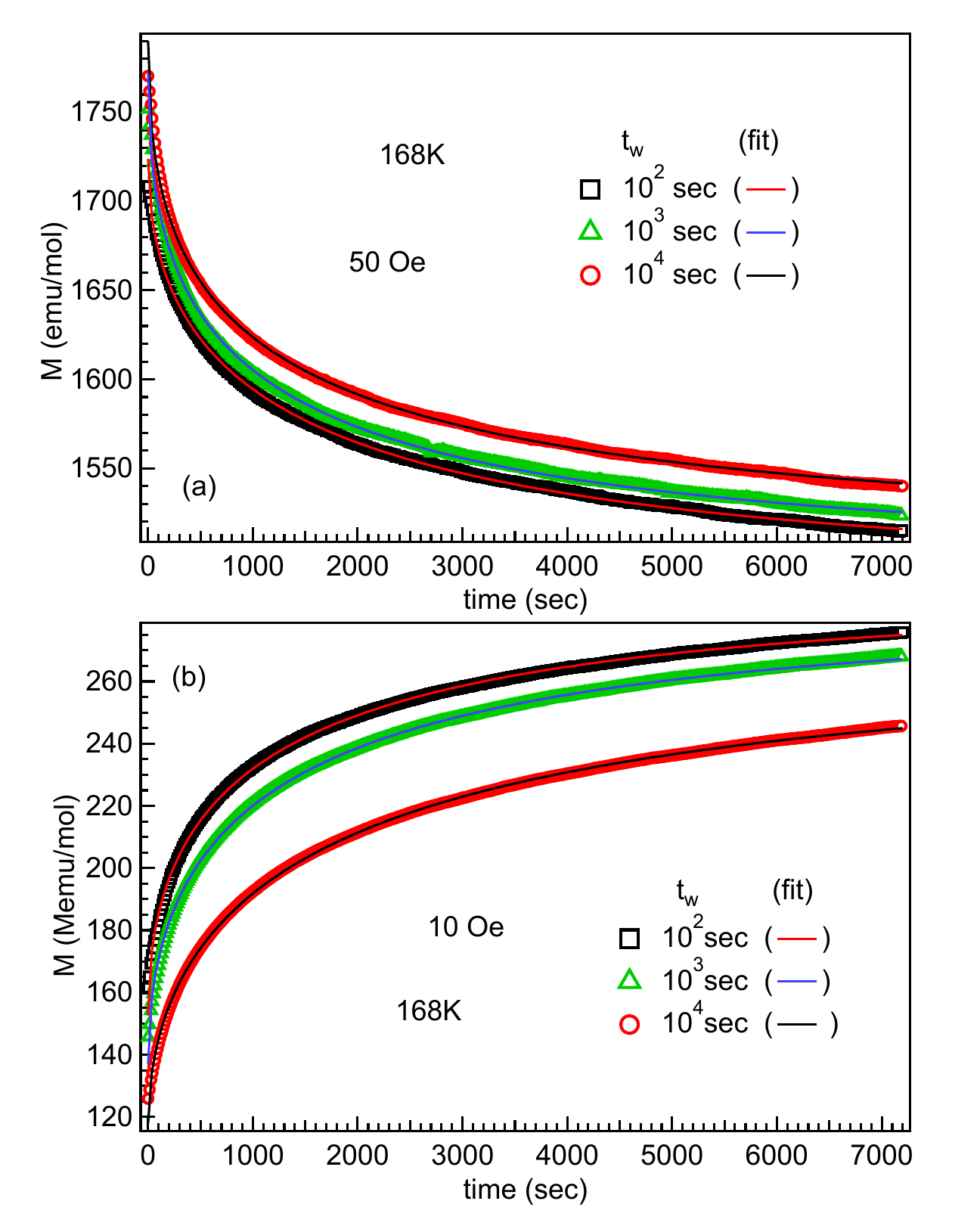}
\caption {(a) The FC IRM and (b) ZFC aging effect data at 168~K using 50~Oe and 10 Oe magnetic fields, respectively, for the different waiting times. } 
\label{Fig9_aging}
\end{figure}

In Figs. \ref{Fig9_aging}(a, b), we present the IRM and aging effect data recorded at 168 K (just below $T_{\rm C}$) for different waiting times under low magnetic fields. To perform the aging effect measurements, the sample was cooled from 300 K to 168 K in zero magnetic field, then kept at 168 K in zero field for a certain time ($t_w$), and then the magnetic field was applied, and the time evolution of the sample’s magnetization was recorded. Each time, the sample was heated to room temperature to erase the prior magnetic history. In IRM, increasing waiting time in the set field enhances magnetization [see Fig. \ref{Fig9_aging}(a)], underscoring relatively slow alignment of the spins with time ($t_w$) and confirming the spin-glass behavior of the sample \cite{Mydosh_Book_93, Kumar_PRB_20}. On the other hand, the magnetization decreases with an increase in $t_w$ (in zero field) during the aging effect measurements [see Fig. \ref{Fig9_aging}(b)], indicating enhanced spin stiffness, i.e., relaxation into deeper energy states, consistent with thermally fluctuating glassy spin dynamics \cite{Mydosh_Book_93, Suzuki_PRB_06, Pakhira_PRB_16}. This behavior is further supported by the shift in the relaxation rate, $S = \frac{1}{H}\frac{dM(t)}{d\log(t)}$ with the waiting time, presented in Fig. 7 of \cite{SI}. The solid curves in Figs. \ref{Fig9_aging}(a, b) represent the best fits obtained using the stretched exponent model, and the fitting parameters are summarized in Table 1 of \cite{SI}.

\begin{figure}
\centering
\includegraphics[width=3.5in]{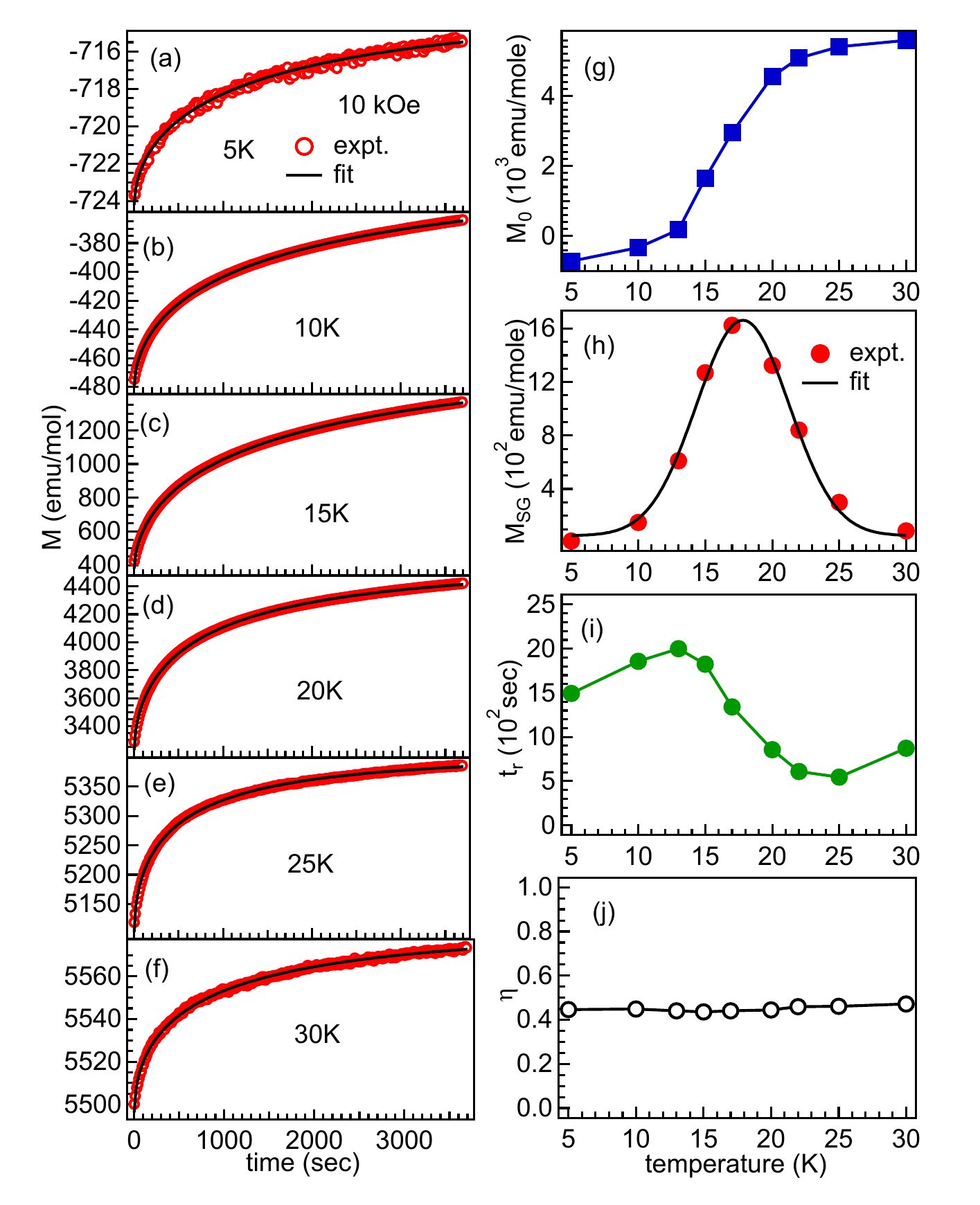}
\caption {(a--f) The ZFC aging effect measurements performed at different temperatures using 10~kOe magnetic field and 600 sec waiting time. The solid black curves represent the best fit using stretched exponent model. The temperature dependence of the (g) ferromagnetic moment, (h) spin-glass moment, (i) relaxation time, and (j) power exponent, $\eta$. The black solid curve in (h) represents the Gaussian fit of the data.} 
\label{Fig10_relax_1T}
\end{figure}

We now discuss the ZFC relaxation (aging effect) measurements performed under higher magnetic fields to elucidate the field-induced changes in the spin dynamics of the sample. In Figs.~\ref{Fig10_relax_1T}(a–f), we present the ZFC relaxation data measured at a 10 kOe field for $t_w = 600$ sec at selected temperatures between 5 and 30~K. A clear logarithmic relaxation of magnetization is observed for up to one hour, even in the presence of a significantly higher field. Note that the sample exhibits negative magnetization at low temperatures [see Figs.~\ref{Fig10_relax_1T}(a, b)], likely due to a nonzero negative remanent field in the SQUID coils, which aligns the magnetic moments upon cooling through $T_{\rm C}$ and subsequently freezes at $T \ll T_{\text{irr}}$. However, this negative moment is not expected to affect the behavior of $M_{\rm SG}$ in the sample, which is governed by the time evolution of the magnetization. Analogous to the IRM measurements performed at 100 Oe, we observe that the relaxation behavior weakens as we move away from a characteristic temperature (say $T_m$) [see Figs.~8(a, b) of~\cite{SI}]. The temperature dependence of the best-fit parameters obtained by fitting the $M(t)$ data using Eq.~\ref{eIRM} is shown in Figs.~\ref{Fig10_relax_1T}(g–j). The $M_0(T)$ curve in Fig.~\ref{Fig10_relax_1T}(g) exhibits behavior similar to the ZFC curve presented in Fig.~\ref{Fig1_ZFC_FC}(e). A prominent peak in the $M_{\rm SG}(T)$ curve is observed at $T_m \sim 17$ K [see Fig.~\ref{Fig10_relax_1T}(h)]. More strikingly, we observe a non-logarithmic change in the characteristic relaxation time, $t_r$ [see Fig.~\ref{Fig10_relax_1T}(i)], in contrast to the behavior seen in the low-field IRM data [see Fig.~\ref{Fig8_IRM}(h)]. This difference indicates that the origin of the field-induced relaxation process is distinct from that of the conventional spin-glass state discussed above.

\begin{figure}
\centering
\includegraphics[width=3.5in]{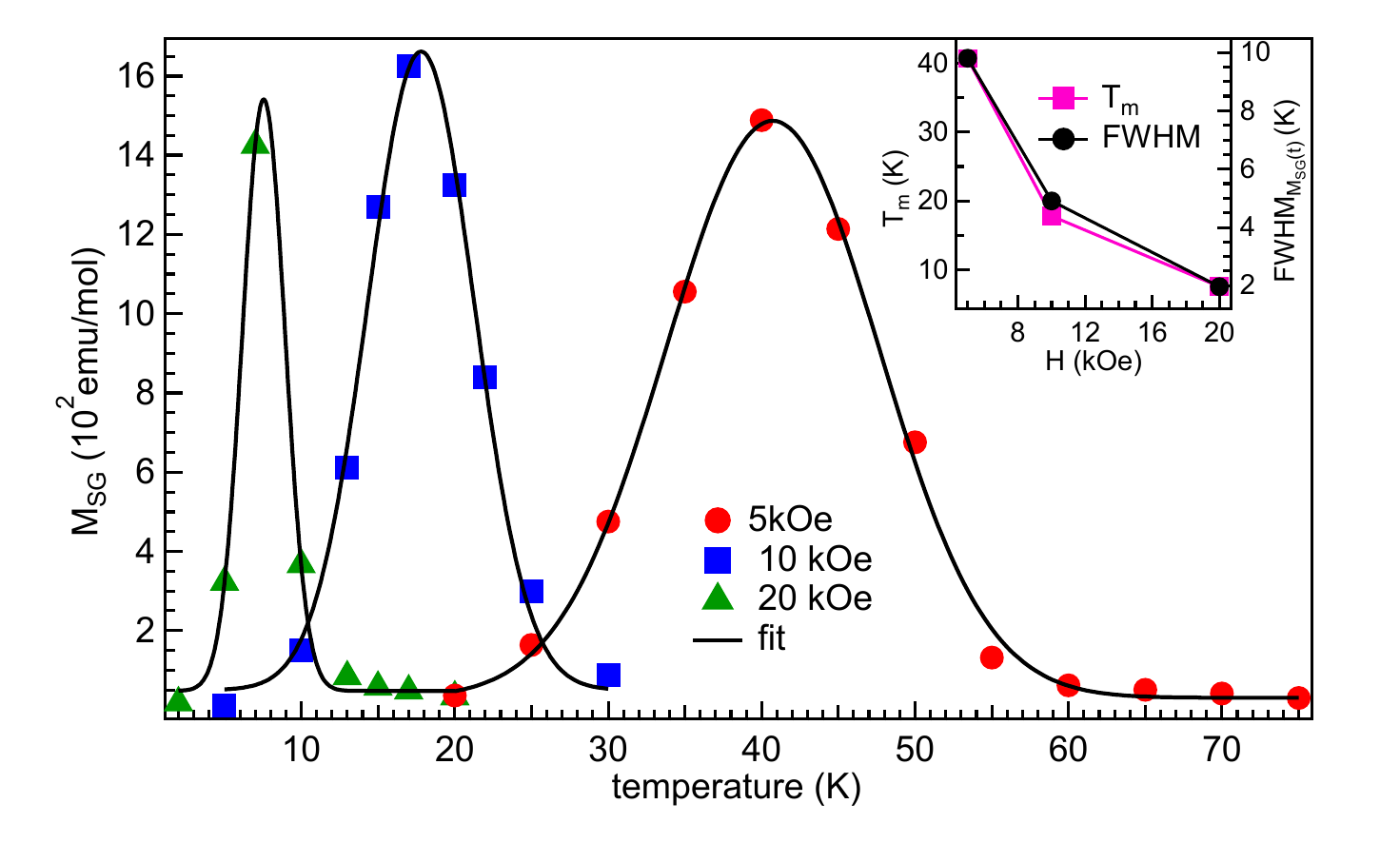}
\caption {The temperature dependence of the spin glass moment (M$_{\rm SG}$) at different fields. The solid black curves represent the Gaussian fit. The inset shows the field dependence of the peak position (on the left axis) and the FWHM (on the right axis) of the M$_{\rm SG}$(T) curves.} 
\label{Fig11_MSG}
\end{figure}

The presence of finite M$_0$ and  M$_{\rm SG}$  in the low-temperature regime for  H = 10~kOe shows that the coexisting region of the glassy and FM components in the sample can be tuned by both the temperature and the external magnetic field. To further clarify this, in Fig.~\ref{Fig11_MSG}, we show the spin-glass moment, M$_{\rm SG}$, as a function of temperature for different magnetic fields, extracted from the ZFC relaxation data [see Figs.~9(a–i) and 10(a–i) of \cite{SI} for details]. It can be observed that the temperature regime spanning the most prominent relaxation behavior (T$_m$) shifts to lower values with an increase in the magnetic field. The peak position and full width at half maximum (FWHM) of all the curves, extracted by Gaussian fitting, are shown in the inset of Fig.~\ref{Fig11_MSG}. It is interesting to note that the glassy region not only shifts to lower temperatures but also becomes more localized with an increase in the external magnetic field. A phenomenologically similar temperature and magnetic field dependence of the spin relaxation behavior has been recently reported in a magnetic glass compound Pr$_{0.5}$Ca$_{0.5}$Mn$_{0.975}$Al$_{0.025}$O$_3$, where the relaxation rate first increases up to T$_m$ = 25~K and then decreases with a further increase in temperature for  H = 10~kOe \cite{Pal_PRB_21}. The peak in the relaxation rate also shifts to lower temperatures with an increase in the magnetic field, and this behavior is attributed to the devitrification of the kinetically arrested high-temperature AFM state into the FM configuration \cite{Pal_PRB_21}. However, the observation of a similar relaxation behavior without any first-order AFM \(\leftrightarrow\) FM transition in the present case indicates a different origin for this glassiness in Sm$_7$Pd$_3$. In the same line, Gd$_5$Ge$_4$ exhibits a pronounced relaxation effect between 10--27.5~kOe, i.e., only in the rising region of the virgin magnetization curve, at 5~K due to the field-induced AFM--FM transformation \cite{Chattopadhyay_PRB_04}. However, no spin relaxation is observed in the compound either below (in the AFM state) or well above (deep in the FM state) a critical field, $H_{\rm cri}$ \cite{Chattopadhyay_PRB_04}. A partial kinetic arrest of the high-field FM state gives rise to strong spin relaxation in Nd$_7$Rh$_3$ below 10~K \cite{Sengupta_PRB_06}, further suggesting a similar relaxation behavior, albeit with a different origin, in Sm$_7$Pd$_3$. It is important to note that the field-induced giant relaxation observed in all these compounds originates from the metastable phase-separated AFM \(\leftrightarrow\) FM regions formed due to the partially arrested first-order magneto-structural transition, and is distinct from the conventional spin-glass behavior that arises from magnetic frustration due to competing FM-AFM interactions, and termed as ``magnetic glass'' state  \cite{Chattopadhyay_PRB_03, Dho_PRB_03, Chattopadhyay_PRB_04, Sengupta_PRB_06, Pal_PRB_21, Pal_PRB_23, Samanta_PRB_14}.

\begin{figure} 
\includegraphics[width=3.5in]{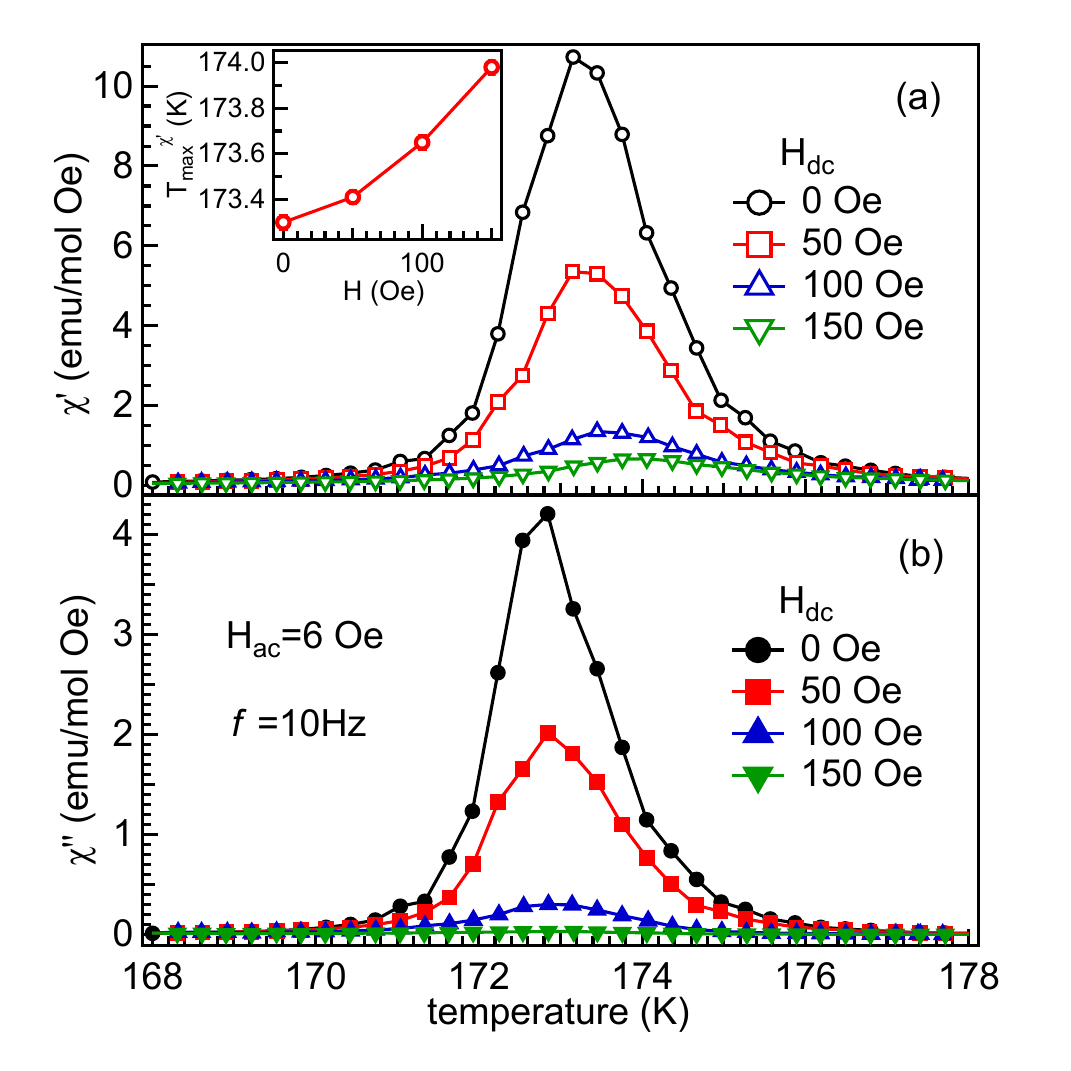}
\caption {The temperature dependent (a) real and (b) imaginary component  of the ac susceptibility at different dc fields using a 6~Oe ac field of 10 Hz frequency. Inset of (a) shows the field dependence of the peak position of $\chi^\prime$(T) curves. }
\label{Fig12_acH}
\end{figure}

In Sm$_7$Pd$_3$, a change in the behavior of the characteristic relaxation time,  t$_r$, in the case of the low- and high-field relaxation measurements discussed above suggests a distinct origin for these two glassy effects. One can further distinguish them by performing the ac susceptibility measurements in the presence of dc magnetic fields, as shown in Figs.~\ref{Fig12_acH}(a, b). If both the low-field (high-temperature) and high-field (low-temperature) glassy behaviors had the same origin, one would expect a shift in the ac susceptibility peak to lower temperatures with the applied field, analogous to the  M$_{\rm SG}$(T, H) curves. However, contrary to this, peaks in both \( \chi^\prime \)- and \( \chi^{\prime\prime} \)-T curves are suppressed and shift to higher temperatures with an increase in the dc field [plotted in the inset of Fig.~\ref{Fig12_acH}(a)]. This confirms that the origin of the glassy effects attributed to the peak of \( \chi^\prime/\chi^{\prime\prime} \)-T curves in Figs.~\ref{Fig6_ac}(a, b) and M$_{\rm SG}$(T) curves in Fig.~\ref{Fig11_MSG} is different. The former arises from the unperturbed competing FM-AFM states in the sample, whereas the latter originates from the relaxation of field-induced metastable states in the sample and more closely resembles the magnetic glass state \cite{Chattopadhyay_PRB_03, Dho_PRB_03, Chattopadhyay_PRB_04, Sengupta_PRB_06, Pal_PRB_21, Pal_PRB_23, Samanta_PRB_14}. However, it may be more precise to refer to it as a ``domain glass" state in the present case to distinguish it from the phase-separated state reported earlier.
 
These experiments provide clear evidence of the unconventional field-induced spin relaxation in Sm$_7$Pd$_3$, even in the absence of a first-order AFM $\leftrightarrow$ FM transition. The observed relaxation in the present case fundamentally originates from the intrinsic magnetic hardness (strong magnetocrystalline anisotropy and hence large H$_{\rm C}$) of the sample. A sufficiently strong magnetic field can align the blocked magnetic moments even within a deeply frozen state. Above a critical magnetic field, H$_{\rm cri}$, the magnetic moments or domain walls begin to orient along the field direction. However, within the polycrystalline material with intrinsically high magnetocrystalline anisotropy, there exists a distribution of domain wall energies required to overcome the pinning effect, causing a nonergodic metastable state to exist in the vicinity of  (T$_{\rm irr}$, H$_{\rm cri}$), thereby leading to slow magnetization relaxation. As the magnetic field is further increased (decreased), the relaxation weakens due to an increase in the number of aligned (frozen) moments. Consequently, only a specific combination of temperature and magnetic field induces these metastable states, which leads to the observed ``domain glass"-like behavior in Sm$_7$Pd$_3$, and can be attributed to the magnetic viscosity of the sample \cite{Sirena_PRB_01, Collocott_PRB_02}.

\begin{figure}
\centering
\includegraphics[width=3.5in]{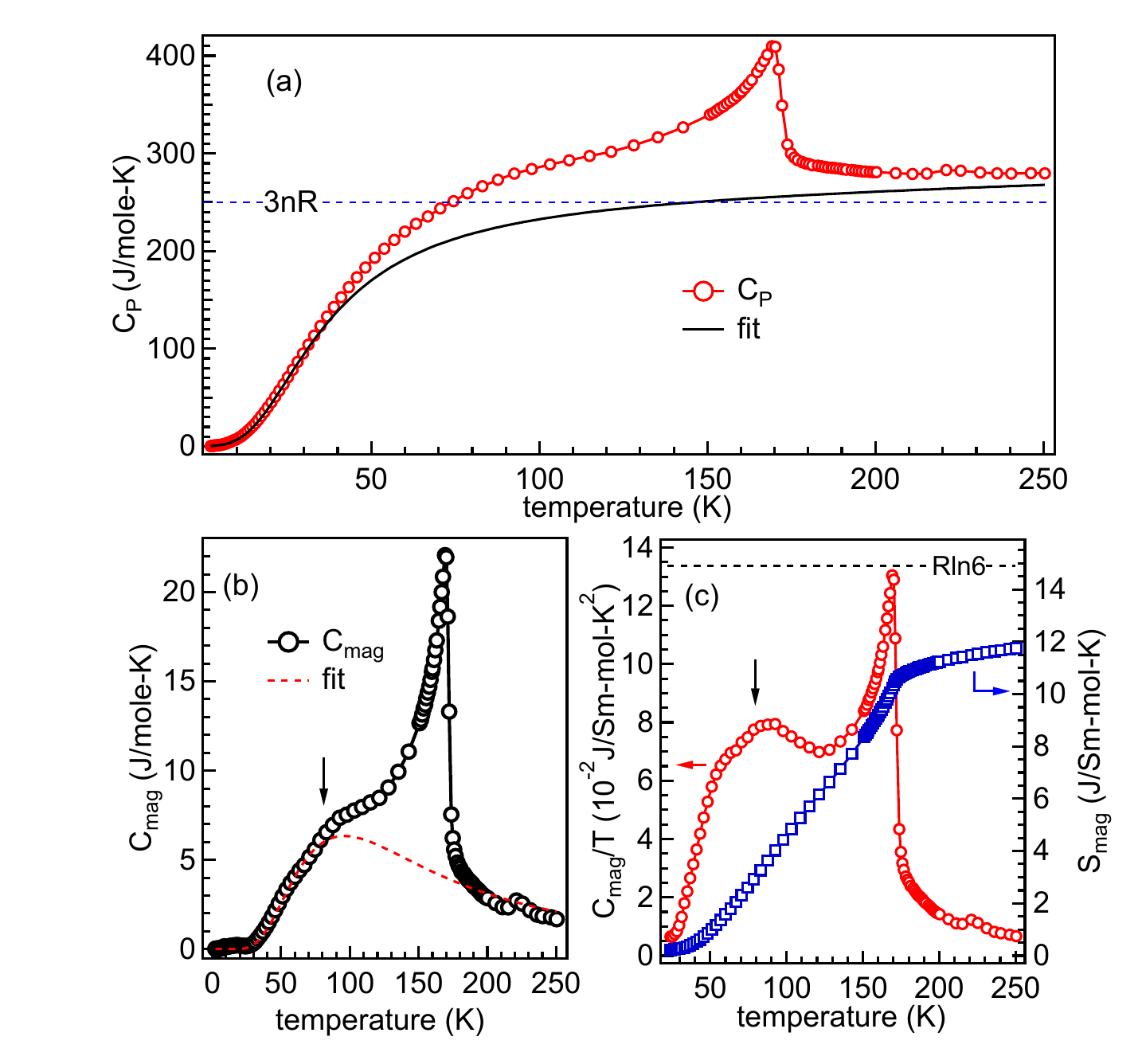}
\caption {(a) The zero-field temperature dependent specific heat of Sm$_7$Pd$_3$.  The dashed line shows the classical Dulong-Petit limit of C$_{\rm P}$. Black solid curve represents the sum of electronic and the lattice contribution (using Debye model), given by eq. \ref{Debye}. (b) The temperature dependent magnetic specific heat, C$_{\rm mag}$, where the arrow represents the Schottky contribution and dashed red curve is the best fit using eq. \ref{schottky}. (c) C$_{\rm mag}$/T vs. T (on left axis) and magnetic entropy (S$_{\rm mag}$) vs. T plot (on right axis). The horizontal dashed line represent the theoretical value of S$_{\rm mag}$ for J = 5/2.}
\label{Fig13_HC}
\end{figure}

In Fig. \ref{Fig13_HC}(a), we show the temperature-dependent zero-field specific heat (C$_{\rm P}$) data of Sm$_7$Pd$_3$. A sharp $\lambda$-like peak at T$_{\rm C}$ is clearly observed, manifesting the presence of long-range magnetic ordering in the sample. Note that the value of specific heat at 250~K is around 280 J/mol-K, which is higher than the classical Dulong-Petit limit of 3$n$R = 249.42 J/mol-K [as shown by the horizontal dashed line in Fig. \ref{Fig13_HC}(a)], where $n$ is the number of atoms per formula unit (ten in the present case), and R is the molar gas constant. This indicates the presence of an additional contribution other than lattice in this temperature range. To estimate the electronic contribution, the low-temperature (2 -- 8~K) data were fitted (not shown) using the equation C$_{\rm P}$ = $\gamma$T + $\beta$T$^3$, where the first and second terms represent the electronic and lattice contributions, respectively, and give $\gamma$ = 91(2) mJ/mol-K$^2$ and $\beta$ = 6.47(5) mJ/mol-K$^4$. The Debye temperature, $\theta_D$, can be estimated from the extracted value of $\beta$ using the following relation \cite{Kittel_book_05}:

\begin{eqnarray} 
\theta_D = \left(\frac{12\pi^4nR}{5\beta}\right)^{1/3}
\end{eqnarray}

We obtained a $\theta_D$ = 145(5)~K and used these values of $\gamma$ and $\theta_D$ to calculate the electronic and lattice specific heat in the entire temperature range using the Debye model as \cite{Kittel_book_05}:

\begin{eqnarray} 
C_P = \gamma T + 9nR\left(\frac{T}{\theta_D}\right)^3 \int_0^{\theta_D/T} \frac{x^4e^x}{(e^x - 1)}dx, 
\label{Debye}
\end{eqnarray}

shown by the solid black curve in Fig. \ref{Fig13_HC}(a). To estimate the magnetic contribution to the specific heat (C$_{\rm mag}$), we subtracted the electronic and lattice specific heat data from the C$_{\rm P}$ curve, and the resulting C$_{\rm mag}$(T) curve is shown in Fig. \ref{Fig13_HC}(b). Interestingly, apart from the $\lambda$ peak at T$_{\rm C}$, we observe a broad Schottky-like anomaly in the 50 -- 120~K temperature range. The crystal field splitting of the six-fold degenerate J = 5/2 ground state of Sm$^{3+}$ is expected to be responsible for this anomaly  \cite{Kumar_PRB_08, Pospisil_PRB_10}. A similar Schottky anomaly is also observed in the specific heat data of Sm-based compounds, Sm$_2$Ni$_2$Sn \cite{Kumar_PRB_08} and SmPd$_2$Al$_3$ \cite{Pospisil_PRB_10}, above their magnetic transition temperatures. We fit this broad peak in the C$_{\rm mag}$ using the following expression \cite{Gopal_book_66}

\begin{eqnarray} 
C_S = \frac {R} {T^2} \left[\frac{\sum_{i} g_i E_i^2 e^{-E_i/T}}{\sum_{i} g_i e^{-E_i/T}} - \left( \frac{\sum_{i} g_i E_i e^{-E_i/T}}{\sum_{i} g_i e^{-E_i/T}} \right)^2 \right],
\label{schottky}
\end{eqnarray}

where E$_i$ and g$_i$ are the energy and degeneracy of the $i$th level. We introduce a three-level splitting scheme with g$_1$ = g$_2$ = g$_3$ = 2 to fit the Schottky contribution; however, the fitting of the C$_{\rm mag}$ data, shown by the dashed red curve in Fig. \ref{Fig13_HC}(b), reveals the presence of a four-fold degenerate state (E$_2$ = E$_3$) lying 255~K above the ground state (E$_1$) doublet, where we fix E$_1$ at 0~K. These low-lying CF-split excited multiplets of Sm$^{3+}$ in Sm$_7$Pd$_3$ are expected to give rise to the temperature-dependent changes in the density of states at the Fermi level, affecting the observed complex magnetic interactions in the sample. Further, the magnetic entropy, S$_{\rm mag}$, is calculated using the relation S$_{\text{mag}}$ = $\int \frac{C_{\text{mag}}}{T} \, dT$. The C$_{\text{mag}}$/T and S$_{\text{mag}}$ are shown on the left and right axes of Fig. \ref{Fig13_HC}(c), respectively. The Schottky anomaly is more clearly visible in the C$_{\text{mag}}$/T, as shown by the downward arrow. The value of S$_{\text{mag}}$ at 250~K is 11.7 J/Sm-mol-K, which is nearly 78\% of the theoretical value of Rln(2J + 1) for J = 5/2, as shown by the dashed horizontal line in Fig. \ref{Fig13_HC}(c). This can be attributed to the underlying crystal field splitting of the J = 5/2 levels and has been observed in several other Sm-based compounds \cite{Ahn_PRB_07}. However, a more reliable estimation of the lattice specific heat using a non-magnetic reference sample is necessary to precisely calculate the C$_{\rm mag}$ and hence S$_{\rm mag}$. No notable change has been observed in the C$_{\rm P}$ curve recorded at H = 10~kOe magnetic field (see Fig. 12 of \cite{SI}) underlying the robustness of this second-order magnetoelastic transition.

\section{\noindent ~Conclusion}

The detailed ac and dc magnetization measurements reveal the coexistence of spin glass, domain glass, and ferromagnetism in the non-centrosymmetric compound Sm$_7$Pd$_3$ below its $T_{\rm C}$ of 173 K.  The dc susceptibility measurements performed in both zero-field-cooled and field-cooled modes exhibit strong irreversibility, which follows the de Almeida-Thouless line, confirming strong magnetic anisotropy in the sample. The temperature-dependent magnetization measurements suggest the presence of finite antiferromagnetic (AFM) coupling in the sample, particularly below $\sim$50 K. However, the field-dependent magnetization data reveal a large coercivity, indicative of long-range ferromagnetic (FM) interactions, along with nearly zero magnetization. This mixed FM-AFM behavior gives rise to spin glass-like dynamics, which is confirmed through detailed ac susceptibility, isothermal remanent magnetization, and aging effect measurements. The analysis of the Cole-Cole plots reveals the distribution of the relaxation time of the frozen spins in the sample. At the same time, a strong field-induced spin relaxation is also observed, which shifts to lower temperatures with an increase in the magnetic field. This phenomenon suggests the presence of domain glass-like behavior in the sample, which is distinct from conventional spin glass systems, and we demonstrate that the intrinsic blocking effect in a hard FM system  due to strong magnetocrystalline anisotropy is sufficient to induce this. 

The analysis of the paramagnetic susceptibility of the sample demonstrates the mixing of ground J = 5/2 state of Sm$^{3+}$ with the first excited J=7/2 multiplets, lying $\sim$ 965~K above. Moreover, the specific heat measurements reveal a $\lambda$-like anomaly at $T_{\rm C}$ associated with long-range magnetic ordering and a pronounced Schottky anomaly between 50 and 120 K due to the splitting of the sixfold degenerate $J = 5/2$ ground state into a ground state doublet, and an excited fourfold-degenerate state situated $\sim$255 K above the doublet, indicating the strong crystal field effect in the sample. Overall, Sm$_7$Pd$_3$ presents an example of the complex magnetic system, where the fundamental description of its magnetism cannot be reduced to classic textbook curriculum. There are still a number of unresolved problems, such as a potential effect of hydrostatic pressure on the devitrification of the magnetic system and its domain alignment; ideally, a single crystal would be highly desired to study magnetic anisotropy. Thus, we encourage the reader to keep an eye out for future publications on the Sm$_7$Pd$_3$ compound.

\subsection{Acknowledgments}

This work was performed at Ames National Laboratory and was supported by the Division of Materials Science and Engineering of the Office of Basic Energy Sciences, Office of Science of the U.S. Department of Energy (DOE). Ames National Laboratory is operated for the U.S. DOE by Iowa State University of Science and Technology under Contract No. DE-AC02-07CH11358.

\end{document}